\documentclass[prl,groupedaddress,amsmath,amssymb,aps,twocolumn]{revtex4-2}

\usepackage{graphicx} 
\usepackage{dcolumn}  
\usepackage{bm}       
\usepackage{hyperref} 

\usepackage{color}

\begin{document}

\title{Anderson transition for light in three dimensions}
\author{Alexey Yamilov}\email{yamilov@mst.edu}
\affiliation{Physics Department, Missouri University of Science \& Technology, Rolla, Missouri 65409}
\author{Hui Cao}\email{hui.cao@yale.edu}
\affiliation{Department of Applied Physics, Yale University, New Haven, Connecticut 06520}
\author{Sergey E. Skipetrov}\email{sergey.skipetrov@lpmmc.cnrs.fr}
\affiliation{Univ. Grenoble Alpes, CNRS, LPMMC, 38000 Grenoble, France}
\date{\today}
\begin{abstract}
\noindent 
We study Anderson transition for light in three dimensions by performing large-scale ab-initio simulations of electromagnetic wave transport in disordered ensembles of conducting spheres. A mobility edge that separates diffusive transport and Anderson localization is identified, revealing a sharp transition from diffusion to localization for light. Critical behavior in the vicinity of the mobility edge is well described by a single-parameter scaling law. The critical exponent is found to be consistent with the value known for the Anderson transition of the orthogonal universality class. Statistical distribution of total transmission at the mobility edge is described without any fit parameter by the diagrammatic perturbation theory originally developed for scalar wave diffusion, but notable deviation from the theory is found when Anderson localization sets in.
\end{abstract}
\maketitle

Anderson localization is a phenomenon of breakdown of quantum or, more generally, wave transport due to interference effects in a disordered medium \cite{Anderson1958,Kramer1993}. The existence of Anderson localization has been suggested for electromagnetic waves in general and light in particular \cite{John1984,Anderson1985}. Localization of light has indeed been observed in low-dimensional systems \cite{Chabanov2000,Lahini2008,Schwartz2007,Riboli2011,Segev2013} but not in three dimensions (3D) \cite{Skipetrov2016} despite numerous attempts \cite{Wiersma1997, Scheffold1999, Wiersma1999, Storzer2006, Beek2012, Sperling2013,  Scheffold2013, Maret2013, Sperling2016}. The current belief is that longitudinal electric fields prevent Anderson localization of light in 3D dielectric disordered media \cite{Skipetrov2014,VanTiggelen2021,Yamilov2023,Micklitz2024}. Recently, ab-initio brute-force numerical solutions of Maxwell equations has led to a discovery of light localization~\cite{Yamilov2023} in 3D fully-disordered ensembles of conducting particles where longitudinal fields either do not exist (in perfect electric conductors) or, at least, are strongly suppressed (in good metals). This finding, however, raises a number of questions: 
(1) Can
the evolution from diffusion to localization of  
light in conducting disordered systems in Ref.~\cite{Yamilov2023} be classified as an Anderson transition?
(2) Is the transition sharp, i.e., does the transition occur at a single frequency that defines a mobility edge? 
(3) Does the transition exhibit universal scaling near the mobility edge as predicted by the standard scaling theory of localization~\cite{Abrahams1979}?
(4) What is the universality class of this transition for electromagnetic waves? 
These are the questions that we seek to answer in the present Letter.

Anderson transition has been studied experimentally in various systems, among which electrical \mbox{(semi-)}conductors \cite{Rosenbaum1980, Paalanen1982} as well as elastic \cite{Hu2008, Cobus2018} and matter \cite{Chabe2008, Kondov2011, Jendr2012, Semeghini2015} waves are the most prominent examples. Theory of Anderson transition provides a good understanding of underlying physics \cite{Evers2008, Wolfle2010} but lacks quantitative accuracy. This gap is successfully filled by numerical approaches \cite{Pichard1981, MacKinnon1981, Slevin1997, Slevin2001, Slevin2014, Delande2014, Rodriguez2010, Rodriguez2011, Pinski2012, Carnio2019}. For light, transitions from diffusion to localization and to photonic band gap regime have been recently studied 
numerically in 3D disordered photonic band gap materials \cite{haberko2020transition, scheffold2022transport}.

A sharp transition between extended and localized states is expected only in an infinitely large system, which is impossible to realize experimentally or in numerical simulations. Such difficulty is circumvented by the finite-size scaling approach \cite{Privman1990} that investigates how the conductance varies with the system size. The mobility edge separating diffusive transport from Anderson localization is crossed as the energy (frequency  $\omega$ for light) is varied. On the diffusion side of the mobility edge, the conductance increases with the system size, while on the localization side it decreases as the system gets larger. Moreover, the localization length diverges at the mobility edge $\omega_c$: $\xi(\omega) \propto|\omega - \omega_c|^{-\nu}$. The critical exponent $\nu$ depends only on the universality class of the transition, and is independent of any details of particular physical systems.

\begin{table*}
\caption{\label{tab:parameters} Parameters of disordered systems simulated and scaling parameters obtained from fitting for the largest $\Delta\omega_{\text{fit}}$.}
\begin{ruledtabular}
\begin{tabular}{ccccccc} 
Radius $r$ (nm) & Filling fraction $f$ & Size $L/\lambda_0$ & Critical exponent $\nu$  & Mobility edge $(\omega_c/c) r$ & $\ln\tilde{g}_c$ & Critical conductance $g_c$\\ 
\hline    
 $25$ & $48\%$ & $3,5,7$ & $1.4\pm 0.4$ & $0.22+0.01$ & $-1.7\pm 0.4$ & $0.40\pm 0.01$\\
 $50$ & $55\%$ & $3,5,7$ & $1.5\pm 0.5$ & $0.44+0.02$ & $-2.4\pm 0.6$ & $0.27\pm 0.01$\\
 $75$ & $61\%$ & $3,5,7$ & $1.6\pm 0.7$ & $0.66+0.03$ & $-2.7\pm 0.7$ & $0.23\pm 0.01$\\
$100$ & $67\%$ & $3,5,7$ & $1.5\pm 0.7$ & $0.93+0.05$ & $-2.8\pm 0.6$ & $0.24\pm 0.01$\\
\end{tabular}
\end{ruledtabular}
\end{table*}

\begin{figure*}
\centering{
\includegraphics[width=7in]{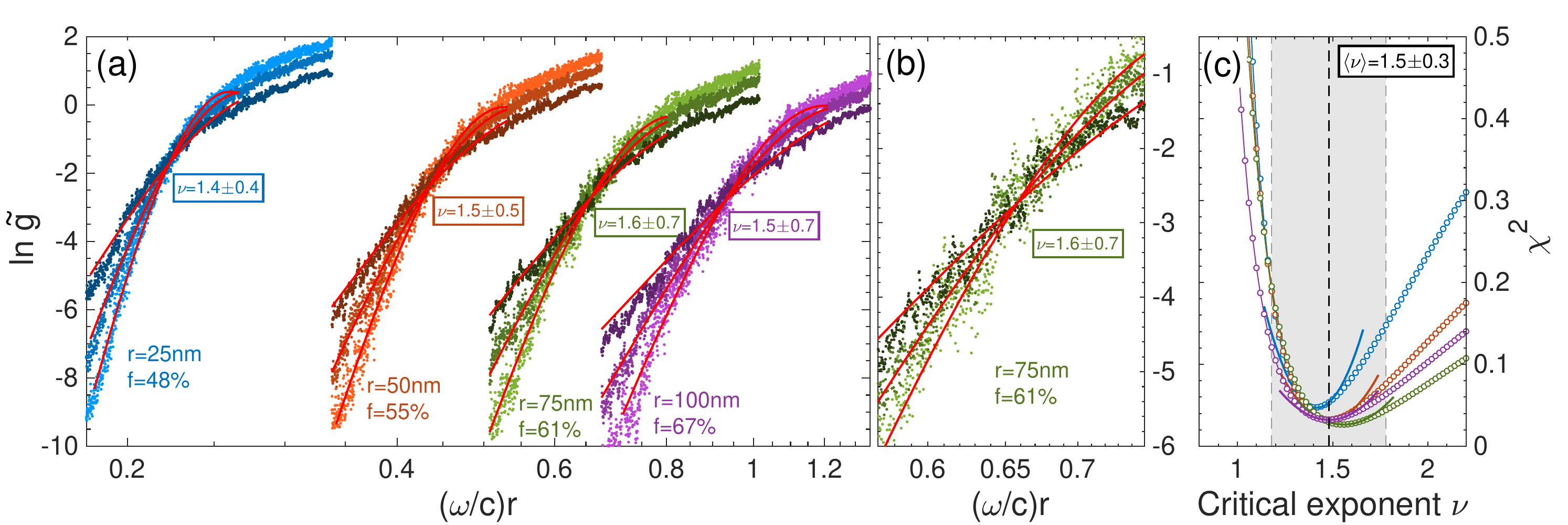}}
\vskip -0.3cm
\caption{\label{fig:scaling} Transition from diffusion to localization in random ensembles of overlapping spheres. (a) Logarithm of the typical conductance, $\ln\tilde{g}$, as a function of normalized frequency $(\omega/c)r$ for different sphere radii and varying system size $L/\lambda_0 = 3$, 5 and 7 (lighter color for larger $L$). For a given $r$, all curves exhibit a crossing. Volume fraction is adjusted to $f = 48$, 55, 61 and 67\% for $r = 25$, 50, 75 and 100 nm (blue, red, green, purple), so that the conductance crossing occurs around $\lambda_0$ = 650 nm for all radii. Solid lines are fits of numerical data to Eq.\ (\ref{eq:scaling_function}), and the critical exponent $\nu$ is given next to each data set. (b) An expanded view of the crossing point for $r=75$ nm. (c) $\chi^2$ statistic versus critical exponent $\nu$ (symbols), and parabolic fits to the numerical data near minima of $\chi^2$ (solid lines). Estimated uncertainty of $\nu$ for each $r$, given in (a), is obtained from the width of fitted parabola in (c). The mean value of $\nu$ and the standard error of the mean are shown in  (c) by the vertical dashed line and the shaded area, respectively.}
\end{figure*}

In this work, we numerically study Anderson transition in 3D fully-disordered systems made of metallic scatterers. Compared to experimental studies, numerical calculations can provide accurate results that are free of experimental artifacts and measurement noises. Moreover, complications from optical absorption may be avoided by simulating perfectly conducting materials. Using a highly efficient, hardware-optimized finite-difference time-domain (FDTD) algorithm~\cite{2024_Flexcompute}, we directly solve Maxwell equations in space and time for 3D random ensembles of perfectly-conducting scatterers. Our numerical results reveal the existence of a mobility edge $\omega_c$ for Anderson transition of light, and confirm the single-parameter scaling of the critical behavior in the vicinity of $\omega_c$. We obtain an estimate for the critical exponent $\nu=1.5\pm 0.3$, which is consistent with previous results for the Anderson transition in the 3D orthogonal universality class. 

To perform the finite-size scaling analysis and to explore the degree of universality of the scaling, we simulate random ensembles of overlapping spheres with four radii $r$. For each sphere size, we simulate light propagation in a $L \times L \times L$ cube. The field transmission coefficient $t_{ab}$ is obtained over a wide frequency range, where $a$ and $b$ denote the input and output modes respectively. By increasing $L$, we track the evolution of total transmission $T_a= \sum_b |t_{ab}|^2$. More details of numerical simulations are given in Supplementary Information \cite{si}. 

In principle, the dimensionless conductance can be obtained by summing over input modes $a$  \cite{1999_van_Rossum,2007_Akkermans_book}: $g = \sum_{a} T_{a}$, but simulating all possible input modes $a$ is too resource consuming. Instead we compute $T_a$ for a single input mode of linearly-polarized plane wave incident normally to the front surface of the scattering system. Ensemble average over disorder realizations gives $\langle g \rangle = (4/5) N \langle T_a \rangle$, where $N = (\omega L/c)^2/2\pi$ is the number of transverse modes  \cite{1999_van_Rossum,2008_Yamilov_PRB}. Thus, $\langle g \rangle$ can be obtained from $\langle T_a \rangle$, but
performing the finite-size scaling of $\langle g \rangle$ would be impractical due to strong fluctuations of $T_a$ from one realization of disorder to another. To circumvent this problem, it is common to work with a typical conductance $\tilde{g}$, which can be, for example, a percentile of the statistical distribution of $g$ or $\exp(\langle \ln g \rangle)$ \cite{Slevin2001,Slevin2014}. The precise choice of $\tilde{g}$ has no importance, although some options turn out to be better adapted to numerical evaluation than the others \cite{Slevin2001}. We choose to average $\ln T_a$ and define
\begin{equation}
    \ln \tilde{g}
    \stackrel{\text{def}}{=}
    \langle \ln [(4/5) N T_a] \rangle
    \label{eq:ln_g_tilde}
\end{equation}

Figure~\ref{fig:scaling}(a) shows $ \ln \tilde{g}$  (dots) versus normalized frequency 
$(\omega/c)r$ 
for four sphere radii $r$ and three system sizes $L$ (Table\ \ref{tab:parameters}). Lighter color corresponds to larger $L$. Statistical averaging is performed over ensembles of $100$, $50$, and $25$ realization for $L/\lambda_0 = 3$, 5 and 7, respectively. For any given $r$, $ \ln \tilde{g}$ increases faster with frequency for larger $L$. Notably, the conductance curves for different $L$ intersect at a single frequency
$\omega_c$. At $\omega < \omega_c$
$\tilde{g}$ decreases with $L$, a signature of localization. For $\omega > \omega_c$, $\tilde{g}$ increases with $L$, consistent with diffusion. This indicates a sharp Anderson transition in the $L\rightarrow\infty$ limit, with the critical frequency $\omega_c$ that can be identified as a mobility edge
\cite{Abrahams1979}.

We now proceed to the quantitative analysis near the mobility edge $\omega_c$. We employ the following numerically robust procedure for the finite-size scaling analysis \cite{Slevin1997,Slevin2014}. We fit $\ln \tilde{g}$ datasets corresponding to a given $r$ in a frequency interval $\pm\Delta\omega_{\text{fit}}$ around the estimated mobility edge to a scaling function 
\begin{equation}
F(\omega, L) = \ln \tilde{g}_c  + B L^{1/\nu} (\omega-\omega_c)+ C L^{2/\nu} (\omega-\omega_c)^2
\label{eq:scaling_function}  
\end{equation}
Values of the five fit parameters $\ln \tilde{g}_c$, $B$, $C$, $\omega_c$ and $\nu$ are obtained by minimizing the $\chi^2$ statistic defined as 
\begin{equation}
\begin{aligned}
\chi^2 &= 
\frac{1}{3}\sum_{i=1}^{3}\frac{1}{N(\Delta\omega_{\text{fit}} r, L_i)}
\sum_{j=1}^{N(\Delta\omega_{\text{fit}}, L_i)} 
\frac{\left[ F(\omega_j, L_i) - \ln \tilde{g}_{ji} \right]^2}{\sigma_{ji}^2}
\end{aligned}
\label{eq:chi2}
\end{equation}
where $N(\Delta\omega_{\text{fit}}, L_i)$ is the number of data points $(\omega_j, \ln\tilde{g}_{ji})$ in $2\Delta\omega_{\text{fit}}$ interval for the system of dimension $L_i$;  $\ln\tilde{g}_{ji}$ is the value of $\ln \tilde{g}$ obtained after ensemble averaging of data at frequency $\omega_j$ for the system of size $ L_i$; $\sigma_{ji}$ is the corresponding standard error of the mean. The fitting results are plotted by solid lines in Fig.\ \ref{fig:scaling}(a), with an expanded view for $r=75$ nm in Fig.\ \ref{fig:scaling}(b). The best-fit parameters $\nu$, $\omega_c$ and $\ln \tilde{g}_c$  are listed in Table\ \ref{tab:parameters}. To obtain their uncertainties, we compute $\chi^2$ as a function of one of them, e.g. $\nu$, with all other  parameters fixed at their best-fit values.
We then perform a parabolic fit of the dependence $\chi^{2}(\nu)$ around its minimum at $\nu = \nu_{\text{best}}$, see Fig.~\ref{fig:scaling}(c), and compute the uncertainty of $\nu$ as  
$\Delta \nu =[2 \partial^2\chi^{2}/\partial\nu^2]^{-1/2}$  at $\nu = \nu_{\text{best}}$ \cite{Bevington1992}. The same analysis is repeated for $\omega_c$ and $\ln \tilde{g}_c$ \cite{si}.

\begin{figure}
\includegraphics[height=2.6in]{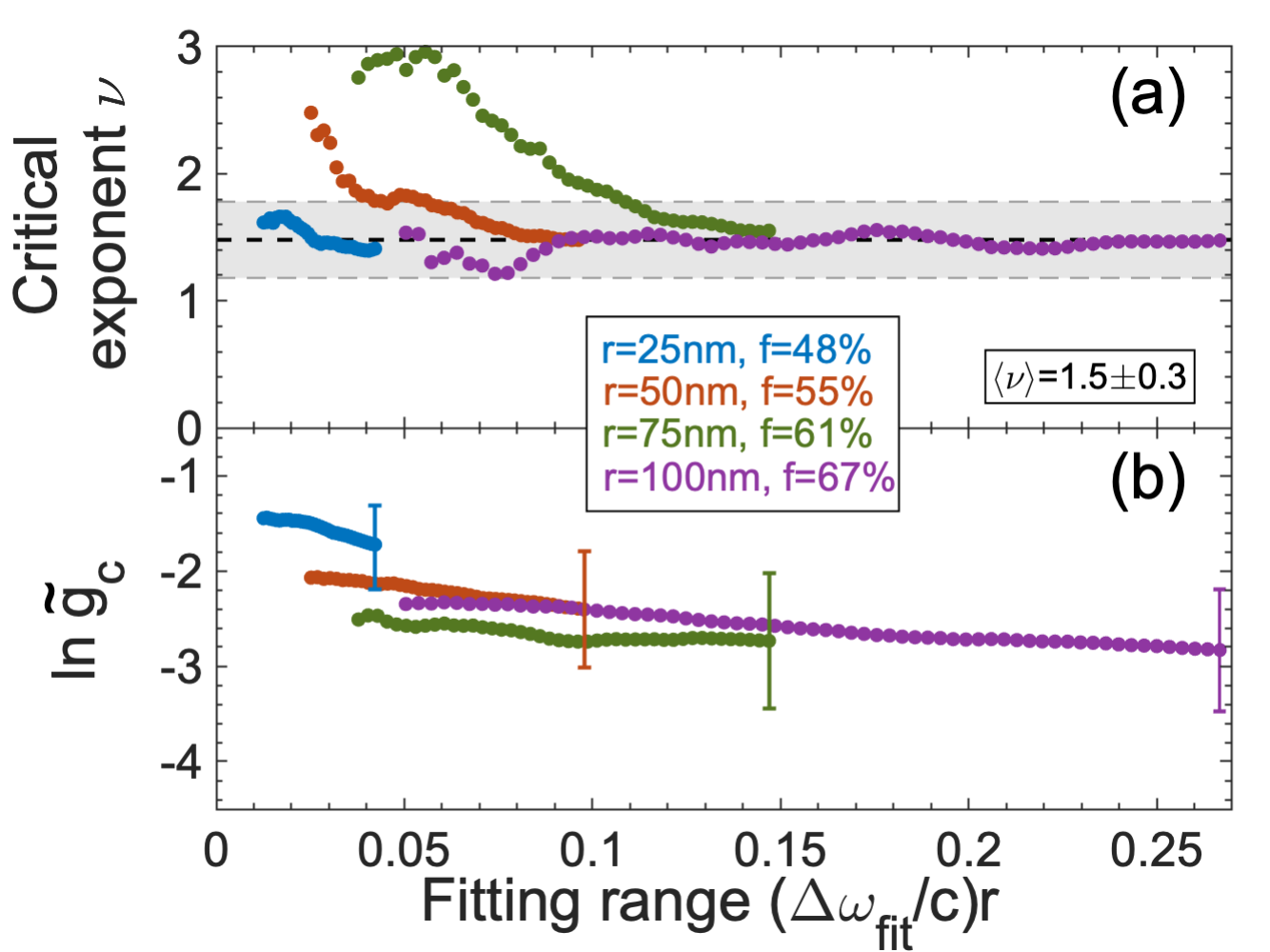}
\vskip -0.3cm
\caption{\label{fig:convergence} Critical exponent and typical conductance at the mobility edge. Values of critical exponent $\nu$ (a) and of the typical conductance at the mobility edge $\ln\tilde{g}_c$ (b) obtained from fitting in the frequency interval $\pm \Delta \omega_{\text{fit}}$ around the estimated mobility edge. In (a), the horizontal dashed line and the shaded gray area show the value of $\nu$ averaged over the four sphere radii at the largest internal $\Delta \omega_{\text{fit}}$ for reach $r$ and its uncertainty, respectively.
In (b), error bars are shown for the largest fitting intervals $\Delta \omega_{\text{fit}}$ \cite{si}.}
\end{figure}

The mobility edge $\omega_c$ follows from the fit
with a rather small uncertainty and  varies very little with $\Delta \omega_{\text{fit}}$ \cite{si}. In contrast, the best-fit value of the critical exponent $\nu$ vary substantially with $\Delta\omega_{\text{fit}}$ but converges robustly towards 1.5 with increasing $\Delta\omega_{\text{fit}}$ for all $r$, see Fig.\ \ref{fig:convergence}(a).
We therefore use the values for the largest $\Delta\omega_{\text{fit}}$ for each $r$ as the best estimates of $\nu$. 
Four systems with different $r$ provide four independent estimates of $\nu$ with corresponding uncertainties listed in Table\ \ref{tab:parameters}. Their mean value and the standard error of the mean provide the final estimate for the critical exponent: $\langle\nu\rangle=1.5\pm 0.3$. This result is consistent with the value $\nu \simeq 1.57$ previously found for Anderson transition of the orthogonal universality class in various systems: Anderson tight-binding model \cite{Slevin1997,Slevin2014}, kicked rotor \cite{Chabe2008,Lemarie2009}, random networks of masses connected by springs \cite{Pinski2012}, elastic waves \cite{Skipetrov2018prb}, light scattering by cold atoms \cite{Skipetrov2018prl}. 

\begin{figure}
\includegraphics[height=2.66in]{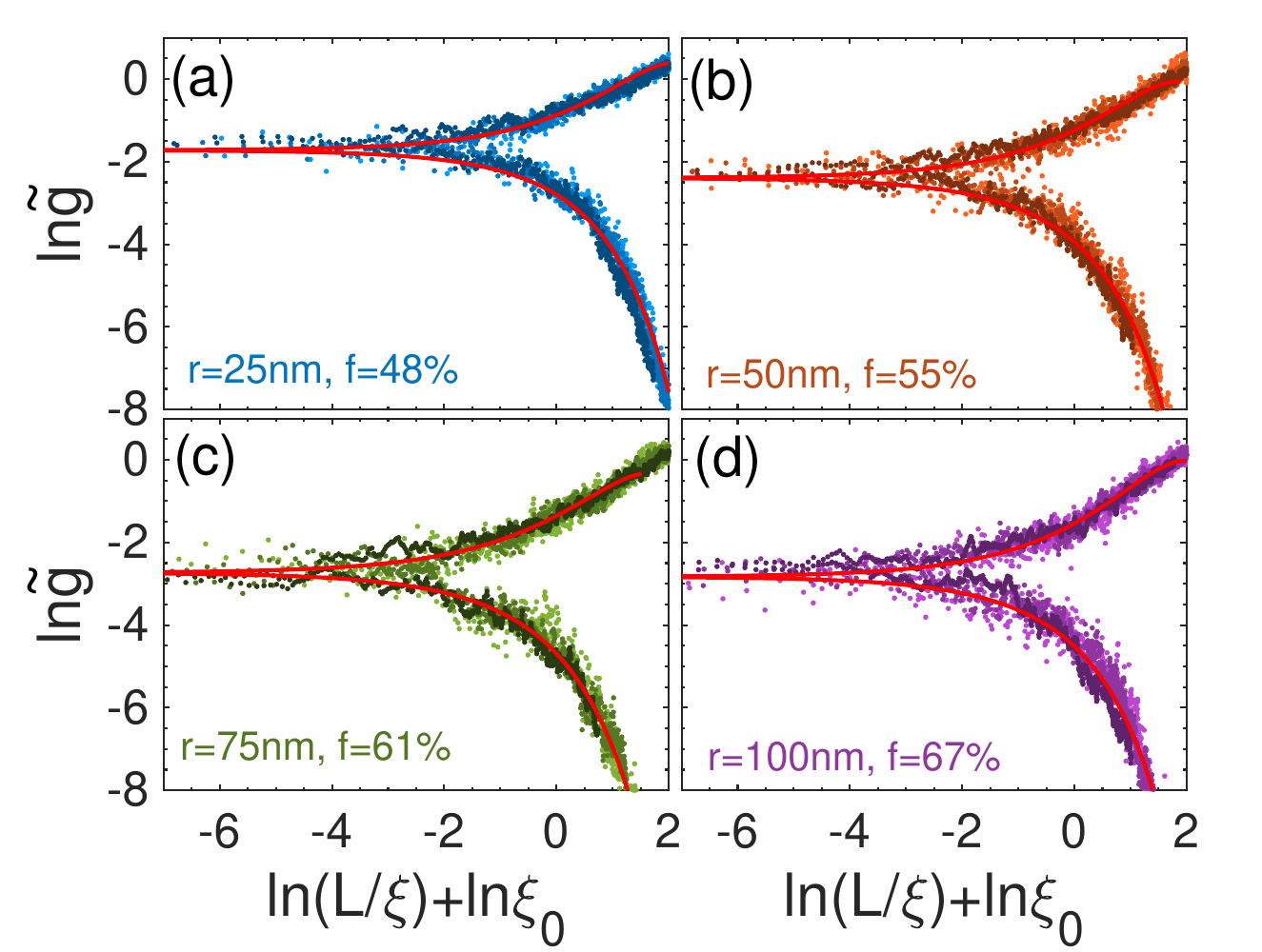}
\vskip -0.3cm
\caption{\label{fig:loc_length} Numerical data from Fig.~\ref{fig:scaling} replotted versus system size $L$ normalized by the localization length $\xi=\xi_0/|\omega-\omega_c|^\nu$. All data collapse on universal curves given by the scaling function (\ref{eq:scaling_function}) shown by solid lines, which confirms the validity of the single-parameter scaling hypothesis.}
\end{figure}

The analysis presented above is based on the hypothesis of single-parameter scaling: $\tilde{g} = \tilde{g}(L/\xi)$, where $\xi$ is the localization length. We check to what extent our numerical results are consistent with this hypothesis. This is done by replotting the data of Fig.\ \ref{fig:scaling} as a function of the ratio $L/\xi$, where $\xi=\xi_0/|\omega-\omega_c|^\nu$, in Fig.\ \ref{fig:loc_length}. The unknown constant $\xi_0$ leads only to a horizontal shift of data points in the chosen logarithmic scale for the horizontal axis. We see that the data points corresponding to different sphere radii [dots in panels (a)--(d)] and different system sizes (colors of different shades) collapse to a single curve given by the scaling function (\ref{eq:scaling_function}) shown by solid lines. This curve has two branches, the top one for diffusion, and the bottom one for localization.  Good agreement between our numerical data and the scaling theory confirms that $\tilde{g}$ can be considered as a function of a single parameter $L/\xi$, and justifies the validity of our analysis \textit{a posteriori}.

Our results of single-parameter scaling with similar values of the critical exponent $\nu$ for all four different sphere sizes $r$ simulated, as well as the proximity of the values of $\nu$ that we find with those from the literature, can be interpreted as a confirmation of the universality of Anderson transition. Namely, the critical behavior is insensitive to microscopic details of disorder as well as to the type of waves and whether they are scalar or vector waves.

The single-parameter scaling shown above relies on the typical conductance $\tilde{g}$ defined in Eq.~(\ref{eq:ln_g_tilde}) as the scaling parameter. Since we define $\tilde{g}$ in terms of the total transmission $T_a$, one may wonder about its relation with the average conductance $\langle g \rangle$, which is believed to be the relevant scaling parameter for the Anderson transition \cite{Evers2008}. It is worthwhile to note that even though we compute $T_a$ and not $g$, we can still obtain the mean $\langle g \rangle = (4/5) N \langle T_a \rangle$ \cite{2007_Akkermans_book}. The value of $\langle g \rangle$ at mobility edge, $g_c$, can be estimated by averaging $\langle g \rangle$ over a narrow frequency interval $\delta\omega$ around $\omega_c$. Using $(\delta\omega/c)r = 0.02$ yields $g_c$ that depends only weakly on the range $\Delta\omega_{\text{fit}}$ of the fit in Fig.\ \ref{fig:scaling}.
For each sphere size $r$, averaging over all $L$'s
for the largest $\Delta\omega_{\text{fit}}$,
yields the best estimate for the critical value $g_c$ given in Table~\ref{tab:parameters}. These values are consistent with the expectation that $g_c$ is on the order of unity \cite{2000_Mirlin}.

\begin{figure}
\includegraphics[height=2.8in]{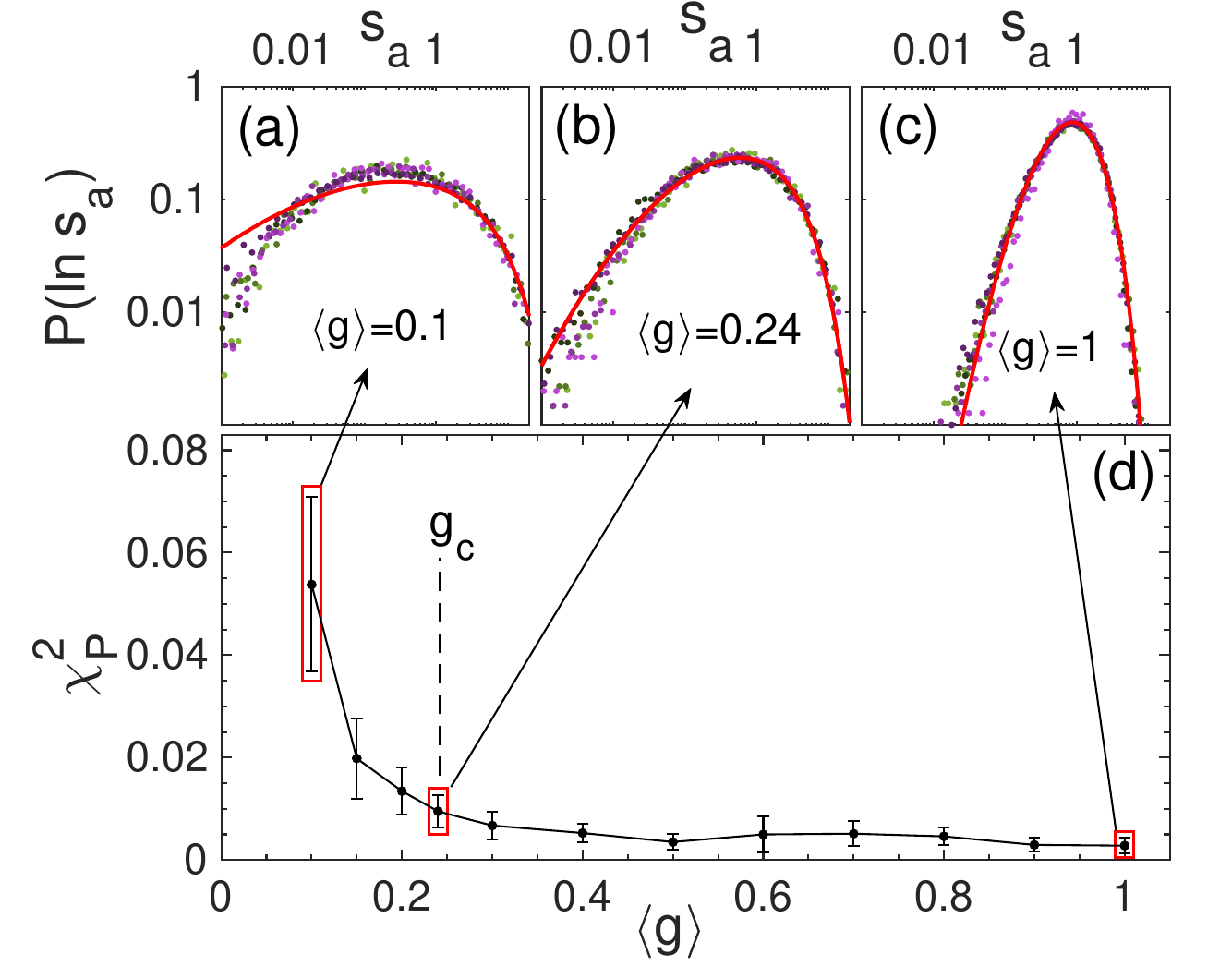}
\vskip -0.3cm
\caption{\label{fig:Pofsa} Statistics of total transmission below, at and above the mobility edge. (a-c) Probability density of the normalized transmission $s_a=T_a/\langle T_a\rangle$ for three values of average conductance $\langle g \rangle$ = $0.1$, $0.24$ ($\simeq g_c$) and $1.0$, obtained by spectral binning of numerical data for $r$ = 75 nm and 100 nm (with nearly identical $g_c$) and all system sizes $L$ in Table~\ref{tab:parameters} (dots). Solid lines are Eq.\ (\ref{eq:Pofsa}) with the actual values of $\langle g\rangle$.
(d) Deviation of Eq.~(\ref{eq:Pofsa}) from the numerical $P(s_a)$ as a function of $\langle g\rangle$. The circles and the error bars represent the mean and standard deviation over different $r$'s and $L$'s, respectively.
}
\end{figure}

To further illustrate on the relation between $\tilde{g}$ in Eq.~(\ref{eq:ln_g_tilde}) and $\langle g \rangle$, we study the full probability distribution of $T_a$.
Figures~\ref{fig:Pofsa}(a--c) show the probability density of the normalized transmission $s_a = T_a/\langle T_a \rangle$ for different $\langle g \rangle$, obtained by sampling the ensemble of disorder realizations within different frequency intervals. It is compared to the prediction of the perturbation diagrammatic theory developed for $\langle g \rangle \gg 1$ \cite{1995_Nieuwenhuizen,1995_Kogan} (red solid lines):
\begin{eqnarray}
    P(s_a) &=& \int_{-i\infty}^{i\infty} \frac{dx}{2\pi i}\exp\left[xs_a-\Phi_{\text{con}}(x)\right] \label{eq:Pofsa}\\
    \Phi_{\text{con}}(x) &=& \langle g\rangle\ln^2\left[\sqrt{1+x/\langle g\rangle}+\sqrt{x/\langle g\rangle} \right] \, .
    \label{eq:Phi}
\end{eqnarray}
Figure~\ref{fig:Pofsa}(a-c) shows the comparison in three regimes: diffuse transport $\langle g \rangle > g_c$, mobility edge $\langle g \rangle = g_c$ and Anderson localization  $\langle g \rangle < g_c$. 
In Fig.~\ref{fig:Pofsa}(d), we quantify the deviation $\chi^2_{P}$ between the numerical and theoretical distributions for a given $\langle g \rangle$ \cite{si}.
For $\langle g \rangle \geq g_c$, the agreement between numerical $P(s_a)$ and Eq.\ (\ref{eq:Pofsa}) is remarkable without any adjustable parameter, with the value of $\langle g \rangle$ that parameterizes $P(s_a)$ calculated directly from $\langle T_a \rangle$ \cite{1999_van_Rossum}. This not only demonstrates the validity of Eq.\ (\ref{eq:Pofsa}) for light near and at the localization transition but also, and more importantly, establishes the equivalence of using $\tilde{g}$ or $\langle g \rangle$ as the relevant scaling parameter for the Anderson transition. Indeed, as far as $P(s_a)$ is parameterized only by $\langle g \rangle$---which is the case according to Fig.\ \ref{fig:Pofsa}---the average of any function of $s_a$, including $\ln\tilde{g} \propto \langle \ln s_a \rangle$, is a function of  $\langle g \rangle$ and thus can be used as a scaling parameter.

Figure\ \ref{fig:Pofsa} also shows that in the regime of Anderson localization where $\langle g \rangle$ becomes significantly smaller than $g_c$, Eq.\ (\ref{eq:Pofsa}) does not hold any more. In particular, the discrepancy between numerical $P(s_a)$ and Eq.\ (\ref{eq:Pofsa}) is significant for $\langle g \rangle = 0.1$. Even if $\langle g \rangle$ is treated as a fitting parameter,  the numerical  $P(s_a)$ cannot be satisfactorily fitted by Eq.\ (\ref{eq:Pofsa})~\cite{si}. Nevertheless, the numerical data for different $r$'s and $L$'s in Fig.~\ref{fig:Pofsa}(a) collapse onto a single curve.    


Finally, we point out that dissipation is absent in perfect electrical conductor media simulated. Any realistic metal has some degree of optical absorption---an aspect that becomes particularly crucial in the context of Anderson localization \cite{Wiersma1997,Scheffold1999,Beek2012}. It would therefore be important to extend the analysis presented in this Letter to disordered media with dissipation. Previous studies have shown that absorption breaks down the single-parameter scaling \cite{Deych2001}, making it necessary to introduce a second scaling parameter. Finite-size scaling in the framework of two-parameter scaling hypothesis may be an interesting extension of our work for Anderson transition of light in realistic 3D systems. 

\acknowledgements
We sincerely thank Prof. Zongfu Yu and Flexcompute Inc. for providing us access to the Tidy3D software for running all the numerical simulations described in this paper.   


\begin{thebibliography}{60}%
\makeatletter
\providecommand \@ifxundefined [1]{%
 \@ifx{#1\undefined}
}%
\providecommand \@ifnum [1]{%
 \ifnum #1\expandafter \@firstoftwo
 \else \expandafter \@secondoftwo
 \fi
}%
\providecommand \@ifx [1]{%
 \ifx #1\expandafter \@firstoftwo
 \else \expandafter \@secondoftwo
 \fi
}%
\providecommand \natexlab [1]{#1}%
\providecommand \enquote  [1]{``#1''}%
\providecommand \bibnamefont  [1]{#1}%
\providecommand \bibfnamefont [1]{#1}%
\providecommand \citenamefont [1]{#1}%
\providecommand \href@noop [0]{\@secondoftwo}%
\providecommand \href [0]{\begingroup \@sanitize@url \@href}%
\providecommand \@href[1]{\@@startlink{#1}\@@href}%
\providecommand \@@href[1]{\endgroup#1\@@endlink}%
\providecommand \@sanitize@url [0]{\catcode `\\12\catcode `\$12\catcode `\&12\catcode `\#12\catcode `\^12\catcode `\_12\catcode `\%12\relax}%
\providecommand \@@startlink[1]{}%
\providecommand \@@endlink[0]{}%
\providecommand \url  [0]{\begingroup\@sanitize@url \@url }%
\providecommand \@url [1]{\endgroup\@href {#1}{\urlprefix }}%
\providecommand \urlprefix  [0]{URL }%
\providecommand \Eprint [0]{\href }%
\providecommand \doibase [0]{https://doi.org/}%
\providecommand \selectlanguage [0]{\@gobble}%
\providecommand \bibinfo  [0]{\@secondoftwo}%
\providecommand \bibfield  [0]{\@secondoftwo}%
\providecommand \translation [1]{[#1]}%
\providecommand \BibitemOpen [0]{}%
\providecommand \bibitemStop [0]{}%
\providecommand \bibitemNoStop [0]{.\EOS\space}%
\providecommand \EOS [0]{\spacefactor3000\relax}%
\providecommand \BibitemShut  [1]{\csname bibitem#1\endcsname}%
\let\auto@bib@innerbib\@empty
\bibitem [{\citenamefont {Anderson}(1958)}]{Anderson1958}%
  \BibitemOpen
  \bibfield  {author} {\bibinfo {author} {\bibfnamefont {P.~W.}\ \bibnamefont {Anderson}},\ }\bibfield  {title} {\bibinfo {title} {Absence of diffusion in certain random lattices},\ }\href {https://doi.org/10.1103/PhysRev.109.1492} {\bibfield  {journal} {\bibinfo  {journal} {Phys. Rev.}\ }\textbf {\bibinfo {volume} {109}},\ \bibinfo {pages} {1492} (\bibinfo {year} {1958})}\BibitemShut {NoStop}%
\bibitem [{\citenamefont {Kramer}\ and\ \citenamefont {MacKinnon}(1993)}]{Kramer1993}%
  \BibitemOpen
  \bibfield  {author} {\bibinfo {author} {\bibfnamefont {B.}~\bibnamefont {Kramer}}\ and\ \bibinfo {author} {\bibfnamefont {A.}~\bibnamefont {MacKinnon}},\ }\bibfield  {title} {\bibinfo {title} {Localization: theory and experiment},\ }\href@noop {} {\bibfield  {journal} {\bibinfo  {journal} {Rep. Prog. Phys.}\ }\textbf {\bibinfo {volume} {56}},\ \bibinfo {pages} {1469} (\bibinfo {year} {1993})}\BibitemShut {NoStop}%
\bibitem [{\citenamefont {John}(1984)}]{John1984}%
  \BibitemOpen
  \bibfield  {author} {\bibinfo {author} {\bibfnamefont {S.}~\bibnamefont {John}},\ }\bibfield  {title} {\bibinfo {title} {Electromagnetic absorption in a disordered medium near a photon mobility edge},\ }\href {https://doi.org/10.1103/PhysRevLett.53.2169} {\bibfield  {journal} {\bibinfo  {journal} {Phys. Rev. Lett.}\ }\textbf {\bibinfo {volume} {53}},\ \bibinfo {pages} {2169} (\bibinfo {year} {1984})}\BibitemShut {NoStop}%
\bibitem [{\citenamefont {Anderson}(1985)}]{Anderson1985}%
  \BibitemOpen
  \bibfield  {author} {\bibinfo {author} {\bibfnamefont {P.~W.}\ \bibnamefont {Anderson}},\ }\bibfield  {title} {\bibinfo {title} {The question of classical localization a theory of white paint?},\ }\href {https://doi.org/10.1080/13642818508240619} {\bibfield  {journal} {\bibinfo  {journal} {Phil. Mag. B}\ }\textbf {\bibinfo {volume} {52}},\ \bibinfo {pages} {505} (\bibinfo {year} {1985})},\ \Eprint {https://arxiv.org/abs/https://doi.org/10.1080/13642818508240619} {https://doi.org/10.1080/13642818508240619} \BibitemShut {NoStop}%
\bibitem [{\citenamefont {Chabanov}\ \emph {et~al.}(2000)\citenamefont {Chabanov}, \citenamefont {Stoytchev},\ and\ \citenamefont {Genack}}]{Chabanov2000}%
  \BibitemOpen
  \bibfield  {author} {\bibinfo {author} {\bibfnamefont {A.}~\bibnamefont {Chabanov}}, \bibinfo {author} {\bibfnamefont {M.}~\bibnamefont {Stoytchev}},\ and\ \bibinfo {author} {\bibfnamefont {A.}~\bibnamefont {Genack}},\ }\bibfield  {title} {\bibinfo {title} {Statistical signatures of photon localization},\ }\href {https://doi.org/10.1038/35009055} {\bibfield  {journal} {\bibinfo  {journal} {Nature}\ }\textbf {\bibinfo {volume} {404}},\ \bibinfo {pages} {850} (\bibinfo {year} {2000})},\ \Eprint {https://arxiv.org/abs/https://doi.org/10.1038/35009055} {https://doi.org/10.1038/35009055} \BibitemShut {NoStop}%
\bibitem [{\citenamefont {Lahini}\ \emph {et~al.}(2008)\citenamefont {Lahini}, \citenamefont {Avidan}, \citenamefont {Pozzi}, \citenamefont {Sorel}, \citenamefont {Morandotti}, \citenamefont {Christodoulides},\ and\ \citenamefont {Silberberg}}]{Lahini2008}%
  \BibitemOpen
  \bibfield  {author} {\bibinfo {author} {\bibfnamefont {Y.}~\bibnamefont {Lahini}}, \bibinfo {author} {\bibfnamefont {A.}~\bibnamefont {Avidan}}, \bibinfo {author} {\bibfnamefont {F.}~\bibnamefont {Pozzi}}, \bibinfo {author} {\bibfnamefont {M.}~\bibnamefont {Sorel}}, \bibinfo {author} {\bibfnamefont {R.}~\bibnamefont {Morandotti}}, \bibinfo {author} {\bibfnamefont {D.~N.}\ \bibnamefont {Christodoulides}},\ and\ \bibinfo {author} {\bibfnamefont {Y.}~\bibnamefont {Silberberg}},\ }\bibfield  {title} {\bibinfo {title} {Anderson localization and nonlinearity in one-dimensional disordered photonic lattices},\ }\href@noop {} {\bibfield  {journal} {\bibinfo  {journal} {Phys. Rev. Lett.}\ }\textbf {\bibinfo {volume} {100}},\ \bibinfo {pages} {013906} (\bibinfo {year} {2008})}\BibitemShut {NoStop}%
\bibitem [{\citenamefont {Schwartz}\ \emph {et~al.}(2007)\citenamefont {Schwartz}, \citenamefont {Bartal}, \citenamefont {Fishman},\ and\ \citenamefont {Segev}}]{Schwartz2007}%
  \BibitemOpen
  \bibfield  {author} {\bibinfo {author} {\bibfnamefont {T.}~\bibnamefont {Schwartz}}, \bibinfo {author} {\bibfnamefont {G.}~\bibnamefont {Bartal}}, \bibinfo {author} {\bibfnamefont {S.}~\bibnamefont {Fishman}},\ and\ \bibinfo {author} {\bibfnamefont {M.}~\bibnamefont {Segev}},\ }\bibfield  {title} {\bibinfo {title} {Transport and {A}nderson localization in disordered two-dimensional photonic lattices},\ }\href {https://doi.org/10.1038/nature05623} {\bibfield  {journal} {\bibinfo  {journal} {Nature}\ }\textbf {\bibinfo {volume} {446}},\ \bibinfo {pages} {52} (\bibinfo {year} {2007})},\ \Eprint {https://arxiv.org/abs/https://doi.org/10.1038/nature05623} {https://doi.org/10.1038/nature05623} \BibitemShut {NoStop}%
\bibitem [{\citenamefont {Riboli}\ \emph {et~al.}(2011)\citenamefont {Riboli}, \citenamefont {Barthelemy}, \citenamefont {Vignolini}, \citenamefont {Intonti}, \citenamefont {Rossi}, \citenamefont {Combrie},\ and\ \citenamefont {Wiersma}}]{Riboli2011}%
  \BibitemOpen
  \bibfield  {author} {\bibinfo {author} {\bibfnamefont {F.}~\bibnamefont {Riboli}}, \bibinfo {author} {\bibfnamefont {P.}~\bibnamefont {Barthelemy}}, \bibinfo {author} {\bibfnamefont {S.}~\bibnamefont {Vignolini}}, \bibinfo {author} {\bibfnamefont {F.}~\bibnamefont {Intonti}}, \bibinfo {author} {\bibfnamefont {A.~D.}\ \bibnamefont {Rossi}}, \bibinfo {author} {\bibfnamefont {S.}~\bibnamefont {Combrie}},\ and\ \bibinfo {author} {\bibfnamefont {D.~S.}\ \bibnamefont {Wiersma}},\ }\bibfield  {title} {\bibinfo {title} {Anderson localization of near-visible light in two dimensions},\ }\href {https://doi.org/10.1364/OL.36.000127} {\bibfield  {journal} {\bibinfo  {journal} {Opt. Lett.}\ }\textbf {\bibinfo {volume} {36}},\ \bibinfo {pages} {127} (\bibinfo {year} {2011})}\BibitemShut {NoStop}%
\bibitem [{\citenamefont {Segev}\ \emph {et~al.}(2013)\citenamefont {Segev}, \citenamefont {Silberberg},\ and\ \citenamefont {Christodoulides}}]{Segev2013}%
  \BibitemOpen
  \bibfield  {author} {\bibinfo {author} {\bibfnamefont {M.}~\bibnamefont {Segev}}, \bibinfo {author} {\bibfnamefont {Y.}~\bibnamefont {Silberberg}},\ and\ \bibinfo {author} {\bibfnamefont {D.~N.}\ \bibnamefont {Christodoulides}},\ }\bibfield  {title} {\bibinfo {title} {Anderson localization of light},\ }\href {https://doi.org/10.1038/nphoton.2013.30} {\bibfield  {journal} {\bibinfo  {journal} {Nat. Photon.}\ }\textbf {\bibinfo {volume} {7}},\ \bibinfo {pages} {197} (\bibinfo {year} {2013})},\ \Eprint {https://arxiv.org/abs/https://doi.org/10.1038/nphoton.2013.30} {https://doi.org/10.1038/nphoton.2013.30} \BibitemShut {NoStop}%
\bibitem [{\citenamefont {Skipetrov}\ and\ \citenamefont {Page}(2016)}]{Skipetrov2016}%
  \BibitemOpen
  \bibfield  {author} {\bibinfo {author} {\bibfnamefont {S.~E.}\ \bibnamefont {Skipetrov}}\ and\ \bibinfo {author} {\bibfnamefont {J.~H.}\ \bibnamefont {Page}},\ }\bibfield  {title} {\bibinfo {title} {Red light for {A}nderson localization},\ }\href {https://doi.org/10.1088/1367-2630/18/2/021001} {\bibfield  {journal} {\bibinfo  {journal} {New J. Phys. Physics}\ }\textbf {\bibinfo {volume} {18}},\ \bibinfo {pages} {021001} (\bibinfo {year} {2016})}\BibitemShut {NoStop}%
\bibitem [{\citenamefont {Wiersma}\ \emph {et~al.}(1997)\citenamefont {Wiersma}, \citenamefont {Bartolini}, \citenamefont {Lagendijk},\ and\ \citenamefont {Righini}}]{Wiersma1997}%
  \BibitemOpen
  \bibfield  {author} {\bibinfo {author} {\bibfnamefont {D.~S.}\ \bibnamefont {Wiersma}}, \bibinfo {author} {\bibfnamefont {P.}~\bibnamefont {Bartolini}}, \bibinfo {author} {\bibfnamefont {A.}~\bibnamefont {Lagendijk}},\ and\ \bibinfo {author} {\bibfnamefont {R.}~\bibnamefont {Righini}},\ }\bibfield  {title} {\bibinfo {title} {Localization of light in a disordered medium},\ }\href {https://doi.org/10.1038/37757} {\bibfield  {journal} {\bibinfo  {journal} {Nature}\ }\textbf {\bibinfo {volume} {390}},\ \bibinfo {pages} {671} (\bibinfo {year} {1997})}\BibitemShut {NoStop}%
\bibitem [{\citenamefont {Scheffold}\ \emph {et~al.}(1999)\citenamefont {Scheffold}, \citenamefont {Lenke}, \citenamefont {Tweer},\ and\ \citenamefont {Maret}}]{Scheffold1999}%
  \BibitemOpen
  \bibfield  {author} {\bibinfo {author} {\bibfnamefont {F.}~\bibnamefont {Scheffold}}, \bibinfo {author} {\bibfnamefont {R.}~\bibnamefont {Lenke}}, \bibinfo {author} {\bibfnamefont {R.}~\bibnamefont {Tweer}},\ and\ \bibinfo {author} {\bibfnamefont {G.}~\bibnamefont {Maret}},\ }\bibfield  {title} {\bibinfo {title} {Localization or classical diffusion of light?},\ }\href@noop {} {\bibfield  {journal} {\bibinfo  {journal} {Nature}\ }\textbf {\bibinfo {volume} {398}},\ \bibinfo {pages} {206} (\bibinfo {year} {1999})}\BibitemShut {NoStop}%
\bibitem [{\citenamefont {Wiersma}\ \emph {et~al.}(1999)\citenamefont {Wiersma}, \citenamefont {Rivas}, \citenamefont {Bartolini}, \citenamefont {Lagendijk},\ and\ \citenamefont {Righini}}]{Wiersma1999}%
  \BibitemOpen
  \bibfield  {author} {\bibinfo {author} {\bibfnamefont {D.~S.}\ \bibnamefont {Wiersma}}, \bibinfo {author} {\bibfnamefont {J.~G.}\ \bibnamefont {Rivas}}, \bibinfo {author} {\bibfnamefont {P.}~\bibnamefont {Bartolini}}, \bibinfo {author} {\bibfnamefont {A.}~\bibnamefont {Lagendijk}},\ and\ \bibinfo {author} {\bibfnamefont {R.}~\bibnamefont {Righini}},\ }\bibfield  {title} {\bibinfo {title} {Reply: Localization or classical diffusion of light?},\ }\href {https://doi.org/10.1038/18350} {\bibfield  {journal} {\bibinfo  {journal} {Nature}\ }\textbf {\bibinfo {volume} {398}},\ \bibinfo {pages} {207} (\bibinfo {year} {1999})}\BibitemShut {NoStop}%
\bibitem [{\citenamefont {St\"orzer}\ \emph {et~al.}(2006)\citenamefont {St\"orzer}, \citenamefont {Gross}, \citenamefont {Aegerter},\ and\ \citenamefont {Maret}}]{Storzer2006}%
  \BibitemOpen
  \bibfield  {author} {\bibinfo {author} {\bibfnamefont {M.}~\bibnamefont {St\"orzer}}, \bibinfo {author} {\bibfnamefont {P.}~\bibnamefont {Gross}}, \bibinfo {author} {\bibfnamefont {C.~M.}\ \bibnamefont {Aegerter}},\ and\ \bibinfo {author} {\bibfnamefont {G.}~\bibnamefont {Maret}},\ }\bibfield  {title} {\bibinfo {title} {Observation of the critical regime near {A}nderson localization of light},\ }\href {https://doi.org/10.1103/PhysRevLett.96.063904} {\bibfield  {journal} {\bibinfo  {journal} {Phys. Rev. Lett.}\ }\textbf {\bibinfo {volume} {96}},\ \bibinfo {pages} {063904} (\bibinfo {year} {2006})}\BibitemShut {NoStop}%
\bibitem [{\citenamefont {van~der Beek}\ \emph {et~al.}(2012)\citenamefont {van~der Beek}, \citenamefont {Barthelemy}, \citenamefont {Johnson}, \citenamefont {Wiersma},\ and\ \citenamefont {Lagendijk}}]{Beek2012}%
  \BibitemOpen
  \bibfield  {author} {\bibinfo {author} {\bibfnamefont {T.}~\bibnamefont {van~der Beek}}, \bibinfo {author} {\bibfnamefont {P.}~\bibnamefont {Barthelemy}}, \bibinfo {author} {\bibfnamefont {P.~M.}\ \bibnamefont {Johnson}}, \bibinfo {author} {\bibfnamefont {D.~S.}\ \bibnamefont {Wiersma}},\ and\ \bibinfo {author} {\bibfnamefont {A.}~\bibnamefont {Lagendijk}},\ }\bibfield  {title} {\bibinfo {title} {Light transport through disordered layers of dense gallium arsenide submicron particles},\ }\href {https://doi.org/10.1103/PhysRevB.85.115401} {\bibfield  {journal} {\bibinfo  {journal} {Phys. Rev. B}\ }\textbf {\bibinfo {volume} {85}},\ \bibinfo {pages} {115401} (\bibinfo {year} {2012})}\BibitemShut {NoStop}%
\bibitem [{\citenamefont {Sperling}\ \emph {et~al.}(2013)\citenamefont {Sperling}, \citenamefont {B\"{u}hrer}, \citenamefont {Aegerter},\ and\ \citenamefont {Maret}}]{Sperling2013}%
  \BibitemOpen
  \bibfield  {author} {\bibinfo {author} {\bibfnamefont {T.}~\bibnamefont {Sperling}}, \bibinfo {author} {\bibfnamefont {W.}~\bibnamefont {B\"{u}hrer}}, \bibinfo {author} {\bibfnamefont {C.~M.}\ \bibnamefont {Aegerter}},\ and\ \bibinfo {author} {\bibfnamefont {G.}~\bibnamefont {Maret}},\ }\bibfield  {title} {\bibinfo {title} {Direct determination of the transition to localization of light in three dimensions},\ }\href {https://doi.org/10.1038/nphoton.2012.313} {\bibfield  {journal} {\bibinfo  {journal} {Nat. Photon.}\ }\textbf {\bibinfo {volume} {7}},\ \bibinfo {pages} {48} (\bibinfo {year} {2013})}\BibitemShut {NoStop}%
\bibitem [{\citenamefont {Scheffold}\ and\ \citenamefont {Wiersma}(2013)}]{Scheffold2013}%
  \BibitemOpen
  \bibfield  {author} {\bibinfo {author} {\bibfnamefont {F.}~\bibnamefont {Scheffold}}\ and\ \bibinfo {author} {\bibfnamefont {D.}~\bibnamefont {Wiersma}},\ }\bibfield  {title} {\bibinfo {title} {Inelastic scattering puts in question recent claims of {A}nderson localization of light},\ }\href {https://doi.org/10.1038/nphoton.2013.210} {\bibfield  {journal} {\bibinfo  {journal} {Nat. Photon.}\ }\textbf {\bibinfo {volume} {7}},\ \bibinfo {pages} {934} (\bibinfo {year} {2013})}\BibitemShut {NoStop}%
\bibitem [{\citenamefont {Maret}\ \emph {et~al.}(2013)\citenamefont {Maret}, \citenamefont {Sperling}, \citenamefont {B\"{u}hrer}, \citenamefont {Lubatsch}, \citenamefont {Frank},\ and\ \citenamefont {Aegerter}}]{Maret2013}%
  \BibitemOpen
  \bibfield  {author} {\bibinfo {author} {\bibfnamefont {G.}~\bibnamefont {Maret}}, \bibinfo {author} {\bibfnamefont {T.}~\bibnamefont {Sperling}}, \bibinfo {author} {\bibfnamefont {W.}~\bibnamefont {B\"{u}hrer}}, \bibinfo {author} {\bibfnamefont {A.}~\bibnamefont {Lubatsch}}, \bibinfo {author} {\bibfnamefont {R.}~\bibnamefont {Frank}},\ and\ \bibinfo {author} {\bibfnamefont {C.~M.}\ \bibnamefont {Aegerter}},\ }\bibfield  {title} {\bibinfo {title} {Inelastic scattering puts in question recent claims of {A}nderson localization of light},\ }\href {https://doi.org/10.1038/nphoton.2013.281} {\bibfield  {journal} {\bibinfo  {journal} {Nat. Photon.}\ }\textbf {\bibinfo {volume} {7}},\ \bibinfo {pages} {934} (\bibinfo {year} {2013})}\BibitemShut {NoStop}%
\bibitem [{\citenamefont {Sperling}\ \emph {et~al.}(2016)\citenamefont {Sperling}, \citenamefont {Schertel}, \citenamefont {Ackermann}, \citenamefont {Aubry}, \citenamefont {Aegerter},\ and\ \citenamefont {Maret}}]{Sperling2016}%
  \BibitemOpen
  \bibfield  {author} {\bibinfo {author} {\bibfnamefont {T.}~\bibnamefont {Sperling}}, \bibinfo {author} {\bibfnamefont {L.}~\bibnamefont {Schertel}}, \bibinfo {author} {\bibfnamefont {M.}~\bibnamefont {Ackermann}}, \bibinfo {author} {\bibfnamefont {G.~J.}\ \bibnamefont {Aubry}}, \bibinfo {author} {\bibfnamefont {C.~M.}\ \bibnamefont {Aegerter}},\ and\ \bibinfo {author} {\bibfnamefont {G.}~\bibnamefont {Maret}},\ }\bibfield  {title} {\bibinfo {title} {Can 3{D} light localization be reached in `white paint'?},\ }\href {https://doi.org/10.1088/1367-2630/18/1/013039} {\bibfield  {journal} {\bibinfo  {journal} {New J. Phys.}\ }\textbf {\bibinfo {volume} {18}},\ \bibinfo {pages} {013039} (\bibinfo {year} {2016})}\BibitemShut {NoStop}%
\bibitem [{\citenamefont {Skipetrov}\ and\ \citenamefont {Sokolov}(2014)}]{Skipetrov2014}%
  \BibitemOpen
  \bibfield  {author} {\bibinfo {author} {\bibfnamefont {S.~E.}\ \bibnamefont {Skipetrov}}\ and\ \bibinfo {author} {\bibfnamefont {I.~M.}\ \bibnamefont {Sokolov}},\ }\bibfield  {title} {\bibinfo {title} {Absence of {A}nderson localization of light in a random ensemble of point scatterers},\ }\href {https://doi.org/10.1103/PhysRevLett.112.023905} {\bibfield  {journal} {\bibinfo  {journal} {Phys. Rev. Lett.}\ }\textbf {\bibinfo {volume} {112}},\ \bibinfo {pages} {023905} (\bibinfo {year} {2014})}\BibitemShut {NoStop}%
\bibitem [{\citenamefont {van Tiggelen}\ and\ \citenamefont {Skipetrov}(2021)}]{VanTiggelen2021}%
  \BibitemOpen
  \bibfield  {author} {\bibinfo {author} {\bibfnamefont {B.~A.}\ \bibnamefont {van Tiggelen}}\ and\ \bibinfo {author} {\bibfnamefont {S.~E.}\ \bibnamefont {Skipetrov}},\ }\bibfield  {title} {\bibinfo {title} {Longitudinal modes in diffusion and localization of light},\ }\href {https://doi.org/10.1103/PhysRevB.103.174204} {\bibfield  {journal} {\bibinfo  {journal} {Phys. Rev. B}\ }\textbf {\bibinfo {volume} {103}},\ \bibinfo {pages} {174204} (\bibinfo {year} {2021})}\BibitemShut {NoStop}%
\bibitem [{\citenamefont {Yamilov}\ \emph {et~al.}(2023)\citenamefont {Yamilov}, \citenamefont {Skipetrov}, \citenamefont {Hughes}, \citenamefont {Minkov}, \citenamefont {Yu},\ and\ \citenamefont {Cao}}]{Yamilov2023}%
  \BibitemOpen
  \bibfield  {author} {\bibinfo {author} {\bibfnamefont {A.}~\bibnamefont {Yamilov}}, \bibinfo {author} {\bibfnamefont {S.~E.}\ \bibnamefont {Skipetrov}}, \bibinfo {author} {\bibfnamefont {T.~W.}\ \bibnamefont {Hughes}}, \bibinfo {author} {\bibfnamefont {M.}~\bibnamefont {Minkov}}, \bibinfo {author} {\bibfnamefont {Z.}~\bibnamefont {Yu}},\ and\ \bibinfo {author} {\bibfnamefont {H.}~\bibnamefont {Cao}},\ }\bibfield  {title} {\bibinfo {title} {Anderson localization of electromagnetic waves in three dimensions},\ }\href {https://doi.org/10.1038/s41567-023-02091-7} {\bibfield  {journal} {\bibinfo  {journal} {Nat. Phys.}\ }\textbf {\bibinfo {volume} {19}},\ \bibinfo {pages} {1308} (\bibinfo {year} {2023})}\BibitemShut {NoStop}%
\bibitem [{\citenamefont {Micklitz}\ and\ \citenamefont {Altland}(2024)}]{Micklitz2024}%
  \BibitemOpen
  \bibfield  {author} {\bibinfo {author} {\bibfnamefont {T.}~\bibnamefont {Micklitz}}\ and\ \bibinfo {author} {\bibfnamefont {A.}~\bibnamefont {Altland}},\ }\href@noop {} {\bibinfo {title} {Topology obstructing {A}nderson localization of light}} (\bibinfo {year} {2024}),\ \Eprint {https://arxiv.org/abs/2405.13277} {arXiv:2405.13277 [cond-mat.dis-nn]} \BibitemShut {NoStop}%
\bibitem [{\citenamefont {Abrahams}\ \emph {et~al.}(1979)\citenamefont {Abrahams}, \citenamefont {Anderson}, \citenamefont {Licciardello},\ and\ \citenamefont {Ramakrishnan}}]{Abrahams1979}%
  \BibitemOpen
  \bibfield  {author} {\bibinfo {author} {\bibfnamefont {E.}~\bibnamefont {Abrahams}}, \bibinfo {author} {\bibfnamefont {P.~W.}\ \bibnamefont {Anderson}}, \bibinfo {author} {\bibfnamefont {D.~C.}\ \bibnamefont {Licciardello}},\ and\ \bibinfo {author} {\bibfnamefont {T.~V.}\ \bibnamefont {Ramakrishnan}},\ }\bibfield  {title} {\bibinfo {title} {Scaling theory of localization: {A}bsence of quantum diffusion in two dimensions},\ }\href {https://doi.org/10.1103/PhysRevLett.42.673} {\bibfield  {journal} {\bibinfo  {journal} {Phys. Rev. Lett.}\ }\textbf {\bibinfo {volume} {42}},\ \bibinfo {pages} {673} (\bibinfo {year} {1979})}\BibitemShut {NoStop}%
\bibitem [{\citenamefont {Rosenbaum}\ \emph {et~al.}(1980)\citenamefont {Rosenbaum}, \citenamefont {Andres}, \citenamefont {Thomas},\ and\ \citenamefont {Bhatt}}]{Rosenbaum1980}%
  \BibitemOpen
  \bibfield  {author} {\bibinfo {author} {\bibfnamefont {T.~F.}\ \bibnamefont {Rosenbaum}}, \bibinfo {author} {\bibfnamefont {K.}~\bibnamefont {Andres}}, \bibinfo {author} {\bibfnamefont {G.~A.}\ \bibnamefont {Thomas}},\ and\ \bibinfo {author} {\bibfnamefont {R.~N.}\ \bibnamefont {Bhatt}},\ }\bibfield  {title} {\bibinfo {title} {Sharp metal-insulator transition in a random solid},\ }\href {https://doi.org/10.1103/PhysRevLett.45.1723} {\bibfield  {journal} {\bibinfo  {journal} {Phys. Rev. Lett.}\ }\textbf {\bibinfo {volume} {45}},\ \bibinfo {pages} {1723} (\bibinfo {year} {1980})}\BibitemShut {NoStop}%
\bibitem [{\citenamefont {Paalanen}\ \emph {et~al.}(1982)\citenamefont {Paalanen}, \citenamefont {Rosenbaum}, \citenamefont {Thomas},\ and\ \citenamefont {Bhatt}}]{Paalanen1982}%
  \BibitemOpen
  \bibfield  {author} {\bibinfo {author} {\bibfnamefont {M.~A.}\ \bibnamefont {Paalanen}}, \bibinfo {author} {\bibfnamefont {T.~F.}\ \bibnamefont {Rosenbaum}}, \bibinfo {author} {\bibfnamefont {G.~A.}\ \bibnamefont {Thomas}},\ and\ \bibinfo {author} {\bibfnamefont {R.~N.}\ \bibnamefont {Bhatt}},\ }\bibfield  {title} {\bibinfo {title} {Stress tuning of the metal-insulator transition at millikelvin temperatures},\ }\href {https://doi.org/10.1103/PhysRevLett.48.1284} {\bibfield  {journal} {\bibinfo  {journal} {Phys. Rev. Lett.}\ }\textbf {\bibinfo {volume} {48}},\ \bibinfo {pages} {1284} (\bibinfo {year} {1982})}\BibitemShut {NoStop}%
\bibitem [{\citenamefont {Hu}\ \emph {et~al.}(2008)\citenamefont {Hu}, \citenamefont {Strybulevych}, \citenamefont {Page}, \citenamefont {Skipetrov},\ and\ \citenamefont {van Tiggelen}}]{Hu2008}%
  \BibitemOpen
  \bibfield  {author} {\bibinfo {author} {\bibfnamefont {H.}~\bibnamefont {Hu}}, \bibinfo {author} {\bibfnamefont {A.}~\bibnamefont {Strybulevych}}, \bibinfo {author} {\bibfnamefont {J.~H.}\ \bibnamefont {Page}}, \bibinfo {author} {\bibfnamefont {S.~E.}\ \bibnamefont {Skipetrov}},\ and\ \bibinfo {author} {\bibfnamefont {B.~A.}\ \bibnamefont {van Tiggelen}},\ }\bibfield  {title} {\bibinfo {title} {Localization of ultrasound in a three-dimensional elastic network},\ }\href {https://doi.org/10.1038/nphys1101} {\bibfield  {journal} {\bibinfo  {journal} {Nat. Phys.}\ }\textbf {\bibinfo {volume} {4}},\ \bibinfo {pages} {945} (\bibinfo {year} {2008})}\BibitemShut {NoStop}%
\bibitem [{\citenamefont {Cobus}\ \emph {et~al.}(2018)\citenamefont {Cobus}, \citenamefont {Hildebrand}, \citenamefont {Skipetrov}, \citenamefont {van Tiggelen},\ and\ \citenamefont {Page}}]{Cobus2018}%
  \BibitemOpen
  \bibfield  {author} {\bibinfo {author} {\bibfnamefont {L.~A.}\ \bibnamefont {Cobus}}, \bibinfo {author} {\bibfnamefont {W.~K.}\ \bibnamefont {Hildebrand}}, \bibinfo {author} {\bibfnamefont {S.~E.}\ \bibnamefont {Skipetrov}}, \bibinfo {author} {\bibfnamefont {B.~A.}\ \bibnamefont {van Tiggelen}},\ and\ \bibinfo {author} {\bibfnamefont {J.~H.}\ \bibnamefont {Page}},\ }\bibfield  {title} {\bibinfo {title} {Transverse confinement of ultrasound through the {A}nderson transition in three-dimensional mesoglasses},\ }\href {https://doi.org/10.1103/PhysRevB.98.214201} {\bibfield  {journal} {\bibinfo  {journal} {Phys. Rev. B}\ }\textbf {\bibinfo {volume} {98}},\ \bibinfo {pages} {214201} (\bibinfo {year} {2018})}\BibitemShut {NoStop}%
\bibitem [{\citenamefont {Chab\'e}\ \emph {et~al.}(2008)\citenamefont {Chab\'e}, \citenamefont {Lemari\'e}, \citenamefont {Gr\'emaud}, \citenamefont {Delande}, \citenamefont {Szriftgiser},\ and\ \citenamefont {Garreau}}]{Chabe2008}%
  \BibitemOpen
  \bibfield  {author} {\bibinfo {author} {\bibfnamefont {J.}~\bibnamefont {Chab\'e}}, \bibinfo {author} {\bibfnamefont {G.}~\bibnamefont {Lemari\'e}}, \bibinfo {author} {\bibfnamefont {B.}~\bibnamefont {Gr\'emaud}}, \bibinfo {author} {\bibfnamefont {D.}~\bibnamefont {Delande}}, \bibinfo {author} {\bibfnamefont {P.}~\bibnamefont {Szriftgiser}},\ and\ \bibinfo {author} {\bibfnamefont {J.~C.}\ \bibnamefont {Garreau}},\ }\bibfield  {title} {\bibinfo {title} {Experimental observation of the {A}nderson metal-insulator transition with atomic matter waves},\ }\href {https://doi.org/10.1103/PhysRevLett.101.255702} {\bibfield  {journal} {\bibinfo  {journal} {Phys. Rev. Lett.}\ }\textbf {\bibinfo {volume} {101}},\ \bibinfo {pages} {255702} (\bibinfo {year} {2008})}\BibitemShut {NoStop}%
\bibitem [{\citenamefont {Kondov}\ \emph {et~al.}(2011)\citenamefont {Kondov}, \citenamefont {McGehee}, \citenamefont {Zirbel},\ and\ \citenamefont {DeMarco}}]{Kondov2011}%
  \BibitemOpen
  \bibfield  {author} {\bibinfo {author} {\bibfnamefont {S.~S.}\ \bibnamefont {Kondov}}, \bibinfo {author} {\bibfnamefont {W.~R.}\ \bibnamefont {McGehee}}, \bibinfo {author} {\bibfnamefont {J.~J.}\ \bibnamefont {Zirbel}},\ and\ \bibinfo {author} {\bibfnamefont {B.}~\bibnamefont {DeMarco}},\ }\bibfield  {title} {\bibinfo {title} {Three-dimensional {A}nderson localization of ultracold matter},\ }\href {https://doi.org/10.1126/science.1209019} {\bibfield  {journal} {\bibinfo  {journal} {Science}\ }\textbf {\bibinfo {volume} {334}},\ \bibinfo {pages} {66} (\bibinfo {year} {2011})},\ \Eprint {https://arxiv.org/abs/https://www.science.org/doi/pdf/10.1126/science.1209019} {https://www.science.org/doi/pdf/10.1126/science.1209019} \BibitemShut {NoStop}%
\bibitem [{\citenamefont {Jendrzejewski}\ \emph {et~al.}(2012)\citenamefont {Jendrzejewski}, \citenamefont {Bernard}, \citenamefont {M\"{u}ller}, \citenamefont {Cheinet}, \citenamefont {Josse}, \citenamefont {Piraud}, \citenamefont {Pezz\'{e}}, \citenamefont {Sanchez-Palencia}, \citenamefont {Aspect},\ and\ \citenamefont {Bouyer}}]{Jendr2012}%
  \BibitemOpen
  \bibfield  {author} {\bibinfo {author} {\bibfnamefont {F.}~\bibnamefont {Jendrzejewski}}, \bibinfo {author} {\bibfnamefont {A.}~\bibnamefont {Bernard}}, \bibinfo {author} {\bibfnamefont {K.}~\bibnamefont {M\"{u}ller}}, \bibinfo {author} {\bibfnamefont {P.}~\bibnamefont {Cheinet}}, \bibinfo {author} {\bibfnamefont {V.}~\bibnamefont {Josse}}, \bibinfo {author} {\bibfnamefont {M.}~\bibnamefont {Piraud}}, \bibinfo {author} {\bibfnamefont {L.}~\bibnamefont {Pezz\'{e}}}, \bibinfo {author} {\bibfnamefont {L.}~\bibnamefont {Sanchez-Palencia}}, \bibinfo {author} {\bibfnamefont {A.}~\bibnamefont {Aspect}},\ and\ \bibinfo {author} {\bibfnamefont {P.}~\bibnamefont {Bouyer}},\ }\bibfield  {title} {\bibinfo {title} {Three-dimensional localization of ultracold atoms in an optical disordered potential},\ }\href {https://doi.org/10.1038/nphys2256} {\bibfield  {journal} {\bibinfo  {journal} {Nat. Phys.}\ }\textbf {\bibinfo {volume} {8}},\ \bibinfo {pages} {398} (\bibinfo {year} {2012})}\BibitemShut {NoStop}%
\bibitem [{\citenamefont {Semeghini}\ \emph {et~al.}(2015)\citenamefont {Semeghini}, \citenamefont {Landini}, \citenamefont {Castilho}, \citenamefont {Roy}, \citenamefont {Spagnolli}, \citenamefont {Trenkwalder}, \citenamefont {Fattori}, \citenamefont {Inguscio},\ and\ \citenamefont {Modugno}}]{Semeghini2015}%
  \BibitemOpen
  \bibfield  {author} {\bibinfo {author} {\bibfnamefont {G.}~\bibnamefont {Semeghini}}, \bibinfo {author} {\bibfnamefont {M.}~\bibnamefont {Landini}}, \bibinfo {author} {\bibfnamefont {P.}~\bibnamefont {Castilho}}, \bibinfo {author} {\bibfnamefont {S.}~\bibnamefont {Roy}}, \bibinfo {author} {\bibfnamefont {G.}~\bibnamefont {Spagnolli}}, \bibinfo {author} {\bibfnamefont {A.}~\bibnamefont {Trenkwalder}}, \bibinfo {author} {\bibfnamefont {M.}~\bibnamefont {Fattori}}, \bibinfo {author} {\bibfnamefont {M.}~\bibnamefont {Inguscio}},\ and\ \bibinfo {author} {\bibfnamefont {G.}~\bibnamefont {Modugno}},\ }\bibfield  {title} {\bibinfo {title} {Three-dimensional localization of ultracold atoms in an optical disordered potential},\ }\href {https://doi.org/10.1038/nphys3339} {\bibfield  {journal} {\bibinfo  {journal} {Nat. Phys.}\ }\textbf {\bibinfo {volume} {11}},\ \bibinfo {pages} {554} (\bibinfo {year} {2015})}\BibitemShut {NoStop}%
\bibitem [{\citenamefont {Evers}\ and\ \citenamefont {Mirlin}(2008)}]{Evers2008}%
  \BibitemOpen
  \bibfield  {author} {\bibinfo {author} {\bibfnamefont {F.}~\bibnamefont {Evers}}\ and\ \bibinfo {author} {\bibfnamefont {A.~D.}\ \bibnamefont {Mirlin}},\ }\bibfield  {title} {\bibinfo {title} {Anderson transitions},\ }\href {https://doi.org/10.1103/RevModPhys.80.1355} {\bibfield  {journal} {\bibinfo  {journal} {Rev. Mod. Phys.}\ }\textbf {\bibinfo {volume} {80}},\ \bibinfo {pages} {1355} (\bibinfo {year} {2008})}\BibitemShut {NoStop}%
\bibitem [{\citenamefont {W\"{o}lfle}\ and\ \citenamefont {Vollhardt}(2010)}]{Wolfle2010}%
  \BibitemOpen
  \bibfield  {author} {\bibinfo {author} {\bibfnamefont {P.}~\bibnamefont {W\"{o}lfle}}\ and\ \bibinfo {author} {\bibfnamefont {D.}~\bibnamefont {Vollhardt}},\ }\bibfield  {title} {\bibinfo {title} {Self-consistent theory of {A}nderson localization: General formalism and applications},\ }\href {https://doi.org/10.1142/S0217979210064502} {\bibfield  {journal} {\bibinfo  {journal} {International Journal of Modern Physics B}\ }\textbf {\bibinfo {volume} {24}},\ \bibinfo {pages} {1526} (\bibinfo {year} {2010})},\ \Eprint {https://arxiv.org/abs/https://doi.org/10.1142/S0217979210064502} {https://doi.org/10.1142/S0217979210064502} \BibitemShut {NoStop}%
\bibitem [{\citenamefont {Pichard}\ and\ \citenamefont {Sarma}(1981)}]{Pichard1981}%
  \BibitemOpen
  \bibfield  {author} {\bibinfo {author} {\bibfnamefont {J.~L.}\ \bibnamefont {Pichard}}\ and\ \bibinfo {author} {\bibfnamefont {G.}~\bibnamefont {Sarma}},\ }\bibfield  {title} {\bibinfo {title} {Finite size scaling approach to {A}nderson localisation},\ }\href {https://doi.org/10.1088/0022-3719/14/6/003} {\bibfield  {journal} {\bibinfo  {journal} {J. Phys. C: Solid State Phys.}\ }\textbf {\bibinfo {volume} {14}},\ \bibinfo {pages} {L127} (\bibinfo {year} {1981})}\BibitemShut {NoStop}%
\bibitem [{\citenamefont {MacKinnon}\ and\ \citenamefont {Kramer}(1981)}]{MacKinnon1981}%
  \BibitemOpen
  \bibfield  {author} {\bibinfo {author} {\bibfnamefont {A.}~\bibnamefont {MacKinnon}}\ and\ \bibinfo {author} {\bibfnamefont {B.}~\bibnamefont {Kramer}},\ }\bibfield  {title} {\bibinfo {title} {One-parameter scaling of localization length and conductance in disordered systems},\ }\href {https://doi.org/10.1103/PhysRevLett.47.1546} {\bibfield  {journal} {\bibinfo  {journal} {Phys. Rev. Lett.}\ }\textbf {\bibinfo {volume} {47}},\ \bibinfo {pages} {1546} (\bibinfo {year} {1981})}\BibitemShut {NoStop}%
\bibitem [{\citenamefont {Slevin}\ and\ \citenamefont {Ohtsuki}(1997)}]{Slevin1997}%
  \BibitemOpen
  \bibfield  {author} {\bibinfo {author} {\bibfnamefont {K.}~\bibnamefont {Slevin}}\ and\ \bibinfo {author} {\bibfnamefont {T.}~\bibnamefont {Ohtsuki}},\ }\bibfield  {title} {\bibinfo {title} {The {A}nderson transition: Time reversal symmetry and universality},\ }\href {https://doi.org/10.1103/PhysRevLett.78.4083} {\bibfield  {journal} {\bibinfo  {journal} {Phys. Rev. Lett.}\ }\textbf {\bibinfo {volume} {78}},\ \bibinfo {pages} {4083} (\bibinfo {year} {1997})}\BibitemShut {NoStop}%
\bibitem [{\citenamefont {Slevin}\ \emph {et~al.}(2001)\citenamefont {Slevin}, \citenamefont {Marko\ifmmode~\check{s}\else \v{s}\fi{}},\ and\ \citenamefont {Ohtsuki}}]{Slevin2001}%
  \BibitemOpen
  \bibfield  {author} {\bibinfo {author} {\bibfnamefont {K.}~\bibnamefont {Slevin}}, \bibinfo {author} {\bibfnamefont {P.}~\bibnamefont {Marko\ifmmode~\check{s}\else \v{s}\fi{}}},\ and\ \bibinfo {author} {\bibfnamefont {T.}~\bibnamefont {Ohtsuki}},\ }\bibfield  {title} {\bibinfo {title} {Reconciling conductance fluctuations and the scaling theory of localization},\ }\href {https://doi.org/10.1103/PhysRevLett.86.3594} {\bibfield  {journal} {\bibinfo  {journal} {Phys. Rev. Lett.}\ }\textbf {\bibinfo {volume} {86}},\ \bibinfo {pages} {3594} (\bibinfo {year} {2001})}\BibitemShut {NoStop}%
\bibitem [{\citenamefont {Slevin}\ and\ \citenamefont {Ohtsuki}(2014)}]{Slevin2014}%
  \BibitemOpen
  \bibfield  {author} {\bibinfo {author} {\bibfnamefont {K.}~\bibnamefont {Slevin}}\ and\ \bibinfo {author} {\bibfnamefont {T.}~\bibnamefont {Ohtsuki}},\ }\bibfield  {title} {\bibinfo {title} {Critical exponent for the {A}nderson transition in the three-dimensional orthogonal universality class},\ }\href {https://doi.org/10.1088/1367-2630/16/1/015012} {\bibfield  {journal} {\bibinfo  {journal} {New J. Phys.}\ }\textbf {\bibinfo {volume} {16}},\ \bibinfo {pages} {015012} (\bibinfo {year} {2014})}\BibitemShut {NoStop}%
\bibitem [{\citenamefont {Delande}\ and\ \citenamefont {Orso}(2014)}]{Delande2014}%
  \BibitemOpen
  \bibfield  {author} {\bibinfo {author} {\bibfnamefont {D.}~\bibnamefont {Delande}}\ and\ \bibinfo {author} {\bibfnamefont {G.}~\bibnamefont {Orso}},\ }\bibfield  {title} {\bibinfo {title} {Mobility edge for cold atoms in laser speckle potentials},\ }\href {https://doi.org/10.1103/PhysRevLett.113.060601} {\bibfield  {journal} {\bibinfo  {journal} {Phys. Rev. Lett.}\ }\textbf {\bibinfo {volume} {113}},\ \bibinfo {pages} {060601} (\bibinfo {year} {2014})}\BibitemShut {NoStop}%
\bibitem [{\citenamefont {Rodriguez}\ \emph {et~al.}(2010)\citenamefont {Rodriguez}, \citenamefont {Vasquez}, \citenamefont {Slevin},\ and\ \citenamefont {R\"omer}}]{Rodriguez2010}%
  \BibitemOpen
  \bibfield  {author} {\bibinfo {author} {\bibfnamefont {A.}~\bibnamefont {Rodriguez}}, \bibinfo {author} {\bibfnamefont {L.~J.}\ \bibnamefont {Vasquez}}, \bibinfo {author} {\bibfnamefont {K.}~\bibnamefont {Slevin}},\ and\ \bibinfo {author} {\bibfnamefont {R.~A.}\ \bibnamefont {R\"omer}},\ }\bibfield  {title} {\bibinfo {title} {Critical parameters from a generalized multifractal analysis at the {A}nderson transition},\ }\href {https://doi.org/10.1103/PhysRevLett.105.046403} {\bibfield  {journal} {\bibinfo  {journal} {Phys. Rev. Lett.}\ }\textbf {\bibinfo {volume} {105}},\ \bibinfo {pages} {046403} (\bibinfo {year} {2010})}\BibitemShut {NoStop}%
\bibitem [{\citenamefont {Rodriguez}\ \emph {et~al.}(2011)\citenamefont {Rodriguez}, \citenamefont {Vasquez}, \citenamefont {Slevin},\ and\ \citenamefont {R\"omer}}]{Rodriguez2011}%
  \BibitemOpen
  \bibfield  {author} {\bibinfo {author} {\bibfnamefont {A.}~\bibnamefont {Rodriguez}}, \bibinfo {author} {\bibfnamefont {L.~J.}\ \bibnamefont {Vasquez}}, \bibinfo {author} {\bibfnamefont {K.}~\bibnamefont {Slevin}},\ and\ \bibinfo {author} {\bibfnamefont {R.~A.}\ \bibnamefont {R\"omer}},\ }\bibfield  {title} {\bibinfo {title} {Multifractal finite-size scaling and universality at the {A}nderson transition},\ }\href {https://doi.org/10.1103/PhysRevB.84.134209} {\bibfield  {journal} {\bibinfo  {journal} {Phys. Rev. B}\ }\textbf {\bibinfo {volume} {84}},\ \bibinfo {pages} {134209} (\bibinfo {year} {2011})}\BibitemShut {NoStop}%
\bibitem [{\citenamefont {Pinski}\ \emph {et~al.}(2012)\citenamefont {Pinski}, \citenamefont {Schirmacher},\ and\ \citenamefont {R\"omer}}]{Pinski2012}%
  \BibitemOpen
  \bibfield  {author} {\bibinfo {author} {\bibfnamefont {S.~D.}\ \bibnamefont {Pinski}}, \bibinfo {author} {\bibfnamefont {W.}~\bibnamefont {Schirmacher}},\ and\ \bibinfo {author} {\bibfnamefont {R.~A.}\ \bibnamefont {R\"omer}},\ }\bibfield  {title} {\bibinfo {title} {Anderson universality in a model of disordered phonons},\ }\href {https://doi.org/10.1209/0295-5075/97/16007} {\bibfield  {journal} {\bibinfo  {journal} {Europhys. Lett.}\ }\textbf {\bibinfo {volume} {97}},\ \bibinfo {pages} {16007} (\bibinfo {year} {2012})}\BibitemShut {NoStop}%
\bibitem [{\citenamefont {Carnio}\ \emph {et~al.}(2019)\citenamefont {Carnio}, \citenamefont {Hine},\ and\ \citenamefont {R\"omer}}]{Carnio2019}%
  \BibitemOpen
  \bibfield  {author} {\bibinfo {author} {\bibfnamefont {E.~G.}\ \bibnamefont {Carnio}}, \bibinfo {author} {\bibfnamefont {N.~D.~M.}\ \bibnamefont {Hine}},\ and\ \bibinfo {author} {\bibfnamefont {R.~A.}\ \bibnamefont {R\"omer}},\ }\bibfield  {title} {\bibinfo {title} {Resolution of the exponent puzzle for the {A}nderson transition in doped semiconductors},\ }\href {https://doi.org/10.1103/PhysRevB.99.081201} {\bibfield  {journal} {\bibinfo  {journal} {Phys. Rev. B}\ }\textbf {\bibinfo {volume} {99}},\ \bibinfo {pages} {081201} (\bibinfo {year} {2019})}\BibitemShut {NoStop}%
\bibitem [{\citenamefont {Haberko}\ \emph {et~al.}(2020)\citenamefont {Haberko}, \citenamefont {Froufe-P{\'e}rez},\ and\ \citenamefont {Scheffold}}]{haberko2020transition}%
  \BibitemOpen
  \bibfield  {author} {\bibinfo {author} {\bibfnamefont {J.}~\bibnamefont {Haberko}}, \bibinfo {author} {\bibfnamefont {L.~S.}\ \bibnamefont {Froufe-P{\'e}rez}},\ and\ \bibinfo {author} {\bibfnamefont {F.}~\bibnamefont {Scheffold}},\ }\bibfield  {title} {\bibinfo {title} {Transition from light diffusion to localization in three-dimensional amorphous dielectric networks near the band edge},\ }\href@noop {} {\bibfield  {journal} {\bibinfo  {journal} {Nat. Comm.}\ }\textbf {\bibinfo {volume} {11}},\ \bibinfo {pages} {4867} (\bibinfo {year} {2020})}\BibitemShut {NoStop}%
\bibitem [{\citenamefont {Scheffold}\ \emph {et~al.}(2022)\citenamefont {Scheffold}, \citenamefont {Haberko}, \citenamefont {Magkiriadou},\ and\ \citenamefont {Froufe-P{\'e}rez}}]{scheffold2022transport}%
  \BibitemOpen
  \bibfield  {author} {\bibinfo {author} {\bibfnamefont {F.}~\bibnamefont {Scheffold}}, \bibinfo {author} {\bibfnamefont {J.}~\bibnamefont {Haberko}}, \bibinfo {author} {\bibfnamefont {S.}~\bibnamefont {Magkiriadou}},\ and\ \bibinfo {author} {\bibfnamefont {L.~S.}\ \bibnamefont {Froufe-P{\'e}rez}},\ }\bibfield  {title} {\bibinfo {title} {Transport through amorphous photonic materials with localization and bandgap regimes},\ }\href@noop {} {\bibfield  {journal} {\bibinfo  {journal} {Phys. Rev. Lett.}\ }\textbf {\bibinfo {volume} {129}},\ \bibinfo {pages} {157402} (\bibinfo {year} {2022})}\BibitemShut {NoStop}%
\bibitem [{\citenamefont {Privman}(1990)}]{Privman1990}%
  \BibitemOpen
  \bibfield  {author} {\bibinfo {author} {\bibfnamefont {V.}~\bibnamefont {Privman}},\ }\href {https://doi.org/10.1142/1011} {\emph {\bibinfo {title} {Finite Size Scaling and Numerical Simulation of Statistical Systems}}}\ (\bibinfo  {publisher} {World Scientific},\ \bibinfo {year} {1990})\BibitemShut {NoStop}%
\bibitem [{202()}]{2024_Flexcompute}%
  \BibitemOpen
  \href@noop {} {\bibinfo {title} {Flexcompute inc.}},\ \bibinfo {note} {{h}ttps://flexcompute.com, Accessed: 2024-06-07}\BibitemShut {NoStop}%
\bibitem [{si()}]{si}%
  \BibitemOpen
  \href@noop {} {\bibinfo {title} {Supplementary information}},\ \bibinfo {howpublished} {\url{URL_will_be_inserted_by_publisher}}\BibitemShut {NoStop}%
\bibitem [{\citenamefont {van Rossum}\ and\ \citenamefont {Nieuwenhuizen}(1999)}]{1999_van_Rossum}%
  \BibitemOpen
  \bibfield  {author} {\bibinfo {author} {\bibfnamefont {M.~C.}\ \bibnamefont {van Rossum}}\ and\ \bibinfo {author} {\bibfnamefont {T.~M.}\ \bibnamefont {Nieuwenhuizen}},\ }\bibfield  {title} {\bibinfo {title} {Multiple scattering of classical waves: microscopy, mesoscopy, and diffusion},\ }\href {https://doi.org/10.1103/RevModPhys.71.313} {\bibfield  {journal} {\bibinfo  {journal} {Rev. Mod. Phys.}\ }\textbf {\bibinfo {volume} {71}},\ \bibinfo {pages} {313} (\bibinfo {year} {1999})}\BibitemShut {NoStop}%
\bibitem [{\citenamefont {Akkermans}\ and\ \citenamefont {Montambaux}(2007)}]{2007_Akkermans_book}%
  \BibitemOpen
  \bibfield  {author} {\bibinfo {author} {\bibfnamefont {E.}~\bibnamefont {Akkermans}}\ and\ \bibinfo {author} {\bibfnamefont {G.}~\bibnamefont {Montambaux}},\ }\href@noop {} {\emph {\bibinfo {title} {Mesoscopic Physics of Electrons and Photons}}}\ (\bibinfo  {publisher} {Cambridge University Press},\ \bibinfo {address} {Cambridge, UK},\ \bibinfo {year} {2007})\BibitemShut {NoStop}%
\bibitem [{\citenamefont {Yamilov}(2008)}]{2008_Yamilov_PRB}%
  \BibitemOpen
  \bibfield  {author} {\bibinfo {author} {\bibfnamefont {A.}~\bibnamefont {Yamilov}},\ }\bibfield  {title} {\bibinfo {title} {Relation between channel and spatial mesoscopic correlations in volume-disordered waveguides},\ }\href {http://link.aps.org/doi/10.1103/PhysRevB.78.045104} {\bibfield  {journal} {\bibinfo  {journal} {Phys.~Rev.~B}\ }\textbf {\bibinfo {volume} {78}},\ \bibinfo {eid} {045104} (\bibinfo {year} {2008})}\BibitemShut {NoStop}%
\bibitem [{\citenamefont {Bevington}\ and\ \citenamefont {Robinson}(1992)}]{Bevington1992}%
  \BibitemOpen
  \bibfield  {author} {\bibinfo {author} {\bibfnamefont {P.}~\bibnamefont {Bevington}}\ and\ \bibinfo {author} {\bibfnamefont {D.}~\bibnamefont {Robinson}},\ }\href@noop {} {\emph {\bibinfo {title} {Data Reduction and Error Analysis for the Physical Sciences}}}\ (\bibinfo  {publisher} {McGraw-Hill},\ \bibinfo {year} {1992})\BibitemShut {NoStop}%
\bibitem [{\citenamefont {Lemari\'e}\ \emph {et~al.}(2009)\citenamefont {Lemari\'e}, \citenamefont {Chab\'e}, \citenamefont {Szriftgiser}, \citenamefont {Garreau}, \citenamefont {Gr\'emaud},\ and\ \citenamefont {Delande}}]{Lemarie2009}%
  \BibitemOpen
  \bibfield  {author} {\bibinfo {author} {\bibfnamefont {G.}~\bibnamefont {Lemari\'e}}, \bibinfo {author} {\bibfnamefont {J.}~\bibnamefont {Chab\'e}}, \bibinfo {author} {\bibfnamefont {P.}~\bibnamefont {Szriftgiser}}, \bibinfo {author} {\bibfnamefont {J.~C.}\ \bibnamefont {Garreau}}, \bibinfo {author} {\bibfnamefont {B.}~\bibnamefont {Gr\'emaud}},\ and\ \bibinfo {author} {\bibfnamefont {D.}~\bibnamefont {Delande}},\ }\bibfield  {title} {\bibinfo {title} {Observation of the {A}nderson metal-insulator transition with atomic matter waves: Theory and experiment},\ }\href {https://doi.org/10.1103/PhysRevA.80.043626} {\bibfield  {journal} {\bibinfo  {journal} {Phys. Rev. A}\ }\textbf {\bibinfo {volume} {80}},\ \bibinfo {pages} {043626} (\bibinfo {year} {2009})}\BibitemShut {NoStop}%
\bibitem [{\citenamefont {Skipetrov}\ and\ \citenamefont {Beltukov}(2018)}]{Skipetrov2018prb}%
  \BibitemOpen
  \bibfield  {author} {\bibinfo {author} {\bibfnamefont {S.~E.}\ \bibnamefont {Skipetrov}}\ and\ \bibinfo {author} {\bibfnamefont {Y.~M.}\ \bibnamefont {Beltukov}},\ }\bibfield  {title} {\bibinfo {title} {Anderson transition for elastic waves in three dimensions},\ }\href {https://doi.org/10.1103/PhysRevB.98.064206} {\bibfield  {journal} {\bibinfo  {journal} {Phys. Rev. B}\ }\textbf {\bibinfo {volume} {98}},\ \bibinfo {pages} {064206} (\bibinfo {year} {2018})}\BibitemShut {NoStop}%
\bibitem [{\citenamefont {Skipetrov}(2018)}]{Skipetrov2018prl}%
  \BibitemOpen
  \bibfield  {author} {\bibinfo {author} {\bibfnamefont {S.~E.}\ \bibnamefont {Skipetrov}},\ }\bibfield  {title} {\bibinfo {title} {Localization transition for light scattering by cold atoms in an external magnetic field},\ }\href {https://doi.org/10.1103/PhysRevLett.121.093601} {\bibfield  {journal} {\bibinfo  {journal} {Phys. Rev. Lett.}\ }\textbf {\bibinfo {volume} {121}},\ \bibinfo {pages} {093601} (\bibinfo {year} {2018})}\BibitemShut {NoStop}%
\bibitem [{\citenamefont {Mirlin}(2000)}]{2000_Mirlin}%
  \BibitemOpen
  \bibfield  {author} {\bibinfo {author} {\bibfnamefont {A.}~\bibnamefont {Mirlin}},\ }\bibfield  {title} {\bibinfo {title} {Statistics of energy levels and eigen-functions in disordered systems},\ }\href {https://doi.org/10.1016/S0370-1573(99)00091-5} {\bibfield  {journal} {\bibinfo  {journal} {Phys. Rep.}\ }\textbf {\bibinfo {volume} {326}},\ \bibinfo {pages} {259} (\bibinfo {year} {2000})}\BibitemShut {NoStop}%
\bibitem [{\citenamefont {Nieuwenhuizen}\ and\ \citenamefont {van Rossum}(1995)}]{1995_Nieuwenhuizen}%
  \BibitemOpen
  \bibfield  {author} {\bibinfo {author} {\bibfnamefont {T.~M.}\ \bibnamefont {Nieuwenhuizen}}\ and\ \bibinfo {author} {\bibfnamefont {M.~C.~W.}\ \bibnamefont {van Rossum}},\ }\bibfield  {title} {\bibinfo {title} {Intensity distributions of waves transmitted through a multiple scattering medium},\ }\href {https://doi.org/10.1103/PhysRevLett.74.2674} {\bibfield  {journal} {\bibinfo  {journal} {Phys. Rev. Lett.}\ }\textbf {\bibinfo {volume} {74}},\ \bibinfo {pages} {2674} (\bibinfo {year} {1995})}\BibitemShut {NoStop}%
\bibitem [{\citenamefont {Kogan}\ and\ \citenamefont {Kaveh}(1995)}]{1995_Kogan}%
  \BibitemOpen
  \bibfield  {author} {\bibinfo {author} {\bibfnamefont {E.}~\bibnamefont {Kogan}}\ and\ \bibinfo {author} {\bibfnamefont {M.}~\bibnamefont {Kaveh}},\ }\bibfield  {title} {\bibinfo {title} {Random-matrix-theory approach to the intensity distributions of waves propagating in a random medium},\ }\href {https://doi.org/10.1103/PhysRevB.52.R3813} {\bibfield  {journal} {\bibinfo  {journal} {Phys. Rev. B}\ }\textbf {\bibinfo {volume} {52}},\ \bibinfo {pages} {R3813} (\bibinfo {year} {1995})}\BibitemShut {NoStop}%
\bibitem [{\citenamefont {Deych}\ \emph {et~al.}(2001)\citenamefont {Deych}, \citenamefont {Yamilov},\ and\ \citenamefont {Lisyansky}}]{Deych2001}%
  \BibitemOpen
  \bibfield  {author} {\bibinfo {author} {\bibfnamefont {L.~I.}\ \bibnamefont {Deych}}, \bibinfo {author} {\bibfnamefont {A.}~\bibnamefont {Yamilov}},\ and\ \bibinfo {author} {\bibfnamefont {A.~A.}\ \bibnamefont {Lisyansky}},\ }\bibfield  {title} {\bibinfo {title} {Scaling in one-dimensional localized absorbing systems},\ }\href {https://doi.org/10.1103/PhysRevB.64.024201} {\bibfield  {journal} {\bibinfo  {journal} {Phys. Rev. B}\ }\textbf {\bibinfo {volume} {64}},\ \bibinfo {pages} {024201} (\bibinfo {year} {2001})}\BibitemShut {NoStop}%
\end{thebibliography}
%

\end{document}